\documentclass[twocolumn,preprintnumbers,prb,amsmath,amssymb,superscriptaddress,floatfix]{revtex4}
\usepackage{graphicx}
\usepackage{bm}
\usepackage{amssymb}
\usepackage{units}

\begin{document}

\title{Control of octahedral rotations in (LaNiO$_3$)$_{n}$/(SrMnO$_3$)$_m$ superlattices}

\author{S.\ J.\ May}
\email{smay@drexel.edu}
\affiliation{
Department of Materials Science and Engineering, Drexel University, Philadelphia, PA 19104\\}
\affiliation{
Materials Science Division, Argonne National Laboratory, Argonne, IL 60439\\}
\author{C.\ R.\ Smith}
\affiliation{
Department of Materials Science and Engineering, Drexel University, Philadelphia, PA 19104\\}
\author{J.-W.\ Kim}
\affiliation{
Advanced Photon Source, Argonne National Laboratory, Argonne, IL 60439\\}
\author{E.\ Karapetrova}
\affiliation{
Advanced Photon Source, Argonne National Laboratory, Argonne, IL 60439\\}
\author{A.\ Bhattacharya}
\affiliation{
Materials Science Division, Argonne National Laboratory, Argonne, IL 60439\\}
\affiliation{
Center for Nanoscale Materials, Argonne National Laboratory, Argonne, IL 60439\\}
\author{P.\ J.\ Ryan}
\affiliation{
Advanced Photon Source, Argonne National Laboratory, Argonne, IL 60439\\}

\date{\today}


\begin{abstract}
Oxygen octahedral rotations have been measured in short-period (LaNiO$_3$)$_n$/(SrMnO$_3$)$_m$ superlattices using synchrotron diffraction.  The in-plane and out-of-plane bond angles and lengths are found to systematically vary with superlattice composition.  Rotations are suppressed in structures with $m>n$, producing a nearly cubic form of LaNiO$_3$.  Large rotations are present in structures with $m<n$, leading to reduced bond angles in SrMnO$_3$.  The metal-oxygen-metal bond lengths decrease as rotations are reduced, in contrast to behavior previously observed in strained, single layer films.  This result demonstrates that superlattice structures can be used to stabilize non-equilibrium octahedral behavior in a manner distinct from epitaxial strain, providing a novel means to engineer the electronic and ferroic properties of oxide heterostructures.
\end{abstract}

\maketitle

The $AB$O$_3$ perovskite oxides exhibit an array of physical properties making them attractive for applications in electronics and energy conversion.\cite{Goodenough04,Pena01,Ramesh08}  Recent work has demonstrated that the functionality of these materials can be further expanded through the formation of superlattices, which can exhibit enhanced properties compared to bulk compounds.  While the majority of work in this field has focused on the electronic and ferroic properties of oxide superlattices,\cite{Ohtomo02,Eckstein03,Lee05,Schlom08,Logvenov09,May09b} the impact of heterointerfaces on the local atomic structure has received less attention.\cite{Bousquet08,Jia09,Rondinelli10,Borisevich10}  Of particular importance are the rotations and distortions of the $B$O$_6$ octahedra, which determine the $B$-O-$B$ bond angles ($\theta$) and $B$-O bond lengths ($d$). As both $\theta$ and $d$ couple to the electronic bandwidth, a quantitative understanding of how octahedral rotations are modified in short-period superlattices is necessary in order to control novel electronic phenomena in oxide heterostructures.

Motivated by the question of what octahedral behavior is stabilized at the interface between two structurally dissimilar perovskites, we have investigated the bond angles in a systematic series of (LaNiO$_3$)$_n$/(SrMnO$_3$)$_2$ superlattices.  In bulk, LaNiO$_3$ (LNO) exhibits robust octahedral rotations leading to $\theta$ = 165.2$^\circ$ along all $<0 0 1>$ directions,\cite{Garcia92} with an $a^-a^-a^-$ rotation pattern.\cite{Glazer72}  In contrast, bulk SrMnO$_3$ (SMO) is a cubic perovskite ($a^0a^0a^0$) lacking octahedral rotations ($\theta = 180$$^\circ$).\cite{Chmaissem01}  Due to the short-period of the superlattices investigated and the geometric constraint requiring that the $B$O$_6$ octahedra maintain corner connectivity, the octahedral rotations remain coherent throughout the superlattices.  We demonstrate that the in-plane and out-of-plane bond angles and lengths can be tuned by altering the superlattice composition [$n$/($n$+$m$)], allowing for the stabilization of octahedral behavior that differs substantially from that found in bulk compounds.

  \begin{figure}
\includegraphics[width=2.3 in]{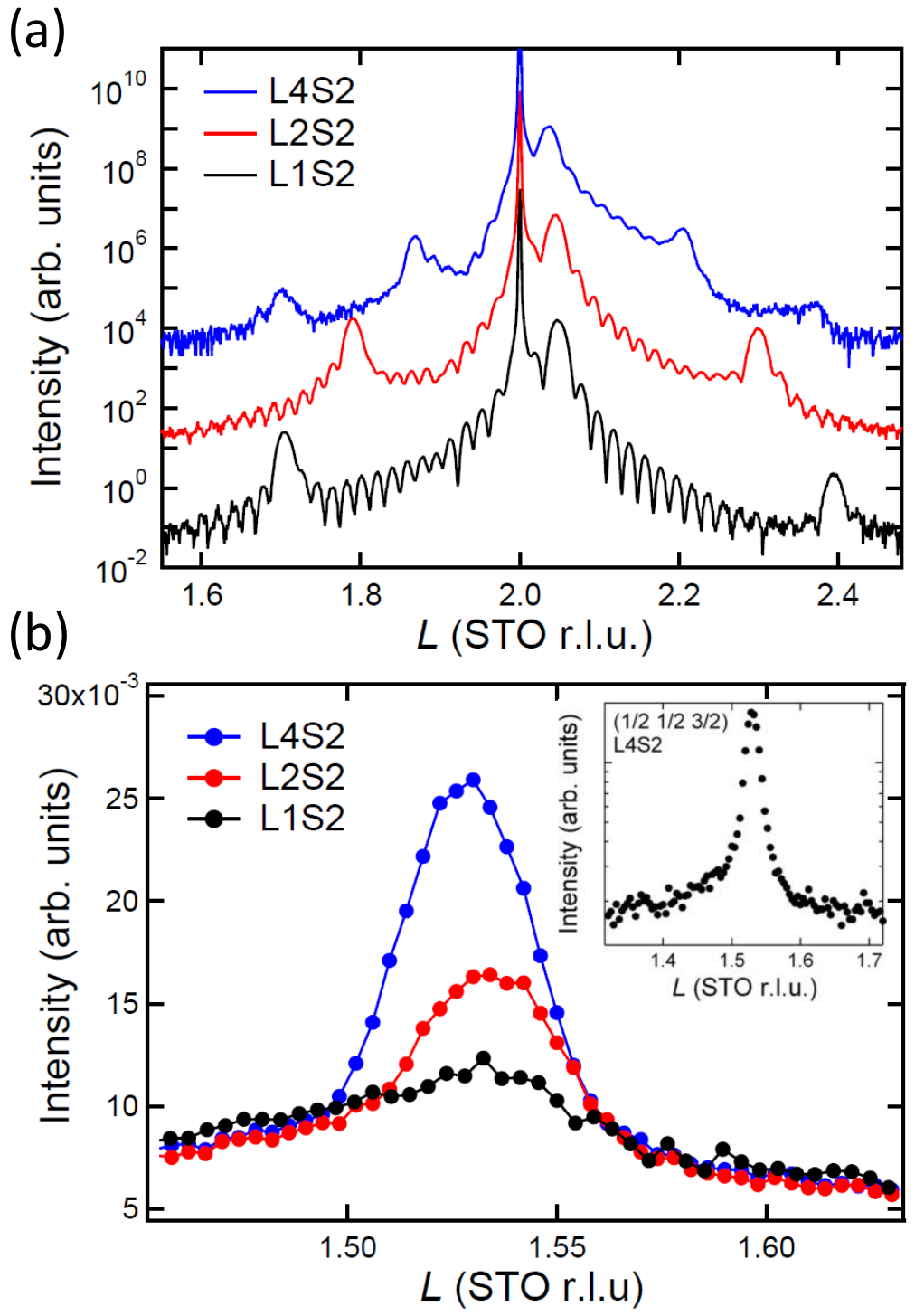}
\caption{(Color online) Scans along (00$L$) in three samples exhibiting distinct satellite peaks due to the coherent superlattice structure (a).  The magnitude of the ($\nicefrac{1}{2}$ $\nicefrac{1}{2}$ $\nicefrac{3}{2}$) peak decreases as the ratio of $m$ to $n$ increases, indicating the magnitude of the octahedral rotations is reduced (b).  Inset of (b), the L4S2 data shown on a log scale; negligible satellite features are observed, consistent with coherent rotational behavior along the $c$-axis.}
\label{fig:tilts}
\end{figure}

The (LNO)$_n$/(SMO)$_m$ superlattices, herein referred to as L$n$S$m$, were deposited on SrTiO$_3$ substrates by ozone-assisted molecular beam epitaxy.\cite{May09}  The sample thicknesses are between 200 and 215 \AA.  Previous structural studies revealed the composition of the superlattices, average $c$-axis parameters, and confirmed that they exhibit minimal interfacial mixing and are highly crystalline.\cite{May09}  Synchrotron diffraction measured along the (00$L$), shown in Fig.~\ref{fig:tilts}(a), provides additional evidence of the superlattice quality.  Satellite peaks are observed for all superlattices, even in the case of the L1S2 sample, confirming that the single LNO layer remains chemically distinct from the SMO layers.  Reciprocal space maps obtained about the superlattice (1 1 3) peaks confirm the epitaxial layers are coherently strained.

The oxygen positions are determined by the measurement and analysis of half-order diffraction peaks arising from the doubling of the unit cell due to octahedral rotations.\cite{May10}  The presence and absence of specific half-order peaks reveals the rotational pattern,\cite{Glazer75, He05, Han10} while the magnitude of the octahedral rotations are determined from the peak intensities.  A systematic survey of the half-order Bragg peaks was carried out on the samples at the Advanced Photon Source on Sectors 6-ID and 33-BM.  All superlattices were found to exhibit Bragg peaks when $h$, $k$, and $l$ are equal to $n$/2, where $n$ is an odd integer, and $h=k\neq l$, $k=l\neq h$, and $h=l\neq k$.  This reciprocal lattice indicates that the presence of an  $a^-a^-c^-$ rotation pattern, consistent with pure LNO films grown on SrTiO$_3$.\cite{May10}  Peaks were not observed at conditions where two Bragg indices were half-order and one was an integer; the lack of such peaks is consistent with the absence of in-phase (for instance $a^+$) rotations.\cite{Glazer75}  Despite all samples exhibiting the same set of peaks, the peak intensities are dependent on the superlattice composition.  This can be seen in Fig.~\ref{fig:tilts}(b), where the ($\nicefrac{1}{2}$ $\nicefrac{1}{2}$ $\nicefrac{3}{2}$) peak is shown to decrease as $n$ is reduced from 4 to 1.  The reduction in peak intensity indicates that the octahedral rotations are suppressed as the number of LNO layers in each superlattice period is reduced.

The half-order intensities were analyzed in two ways to determine the bond angles.  For samples where five or more peaks could be accurately measured, the angles were obtained by minimizing the error between the measured and calculated half-order peak intensities as a function of the octahedral rotations about the [100] and [001] directions, the magnitudes of which are referred to as $\alpha$ and $\gamma$, respectively.\cite{note}  A detailed description of this process is provided in Ref. \onlinecite{May10}; as in that previous work the $A$- and $B$-site cations are assumed to occupy the ideal perovskite lattice positions.  A second method was applied to samples with weak intensities, in which only two or three peaks could be measured.  In these samples the intensities of the measurable peaks, such as the ($\nicefrac{1}{2}$ $\nicefrac{1}{2}$ $\nicefrac{3}{2}$), were compared to those obtained from a superlattice in which $\alpha$ and $\gamma$ were determined using the first method described above.  The rotation angles were determined from the relative peak intensities between the two samples, as the two superlattices are approximately the same thickness.

\begin{figure}
\includegraphics[width=2.9 in]{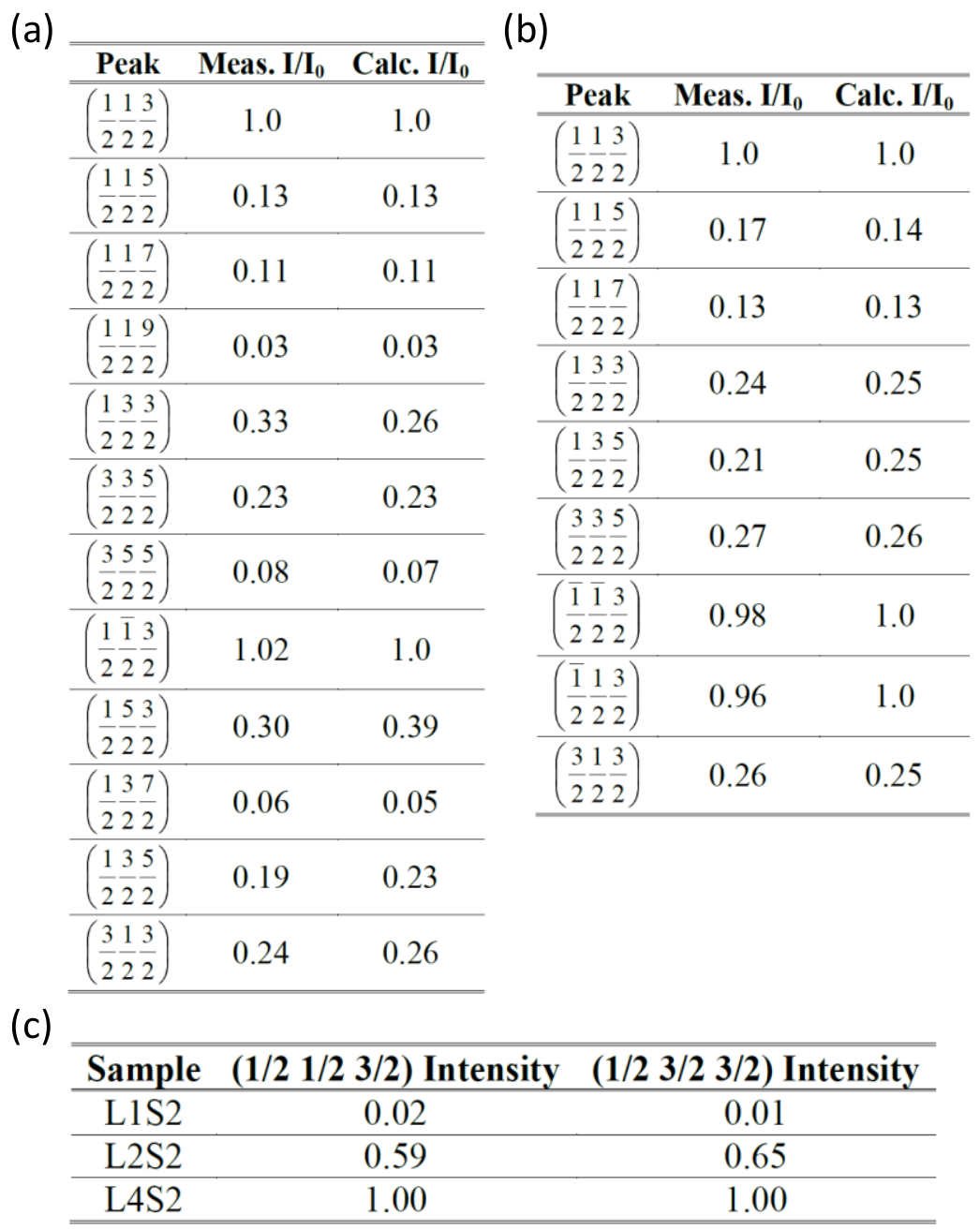}
\caption{The measured and calculated peak intensities for the L4S2 (a) and L2S2 (b) samples.  In (c), the measured intensities of the ($\nicefrac{1}{2}$ $\nicefrac{1}{2}$ $\nicefrac{3}{2}$) and ($\nicefrac{1}{2}$ $\nicefrac{3}{2}$ $\nicefrac{3}{2}$) peaks, normalized by the L4S2 intensities, are given for the three samples.}
\label{fig:table}
\end{figure}

The L4S2 and L2S2 superlattices exhibit robust half-order peaks, allowing for the measurement and analysis of 11 and 7 symmetrically inequivalent Bragg peaks, respectively.  Symmetrically equivalent peaks, such as the ($\nicefrac{1}{2}$ $\nicefrac{1}{2}$ $\nicefrac{3}{2}$) and ($\nicefrac{-1}{2}$ $\nicefrac{1}{2}$ $\nicefrac{3}{2}$), have equal peak intensities confirming that the four orientational domain variants of the $a^-a^-c^-$ structure are present in equal populations.  Fixing the domain volumes to a 1:1:1:1 ratio, the $\alpha$ and $\gamma$ angles were determined by minimizing $\chi^2$ determined from the measured and calculated peak intensities (Fig.~\ref{fig:table}a,b).  Values of $\alpha = 6.6 \pm 0.3^{\circ}$ and $\gamma = 3.5 \pm 1.4^{\circ}$ were obtained for the L4S2 sample, while values of $\alpha = 5.5 \pm 0.2^{\circ}$ and $\gamma = 2.1 \pm 1.0^{\circ}$ were obtained for the L2S2 sample.  The average $\theta$ values were then determined from $\alpha$ and $\gamma$ using:
\begin{equation}
    \theta_{ab} = 180^{\circ} - 2\sqrt{\alpha^2 + \gamma^2},
    \label{eq:one}
\end{equation}
and
\begin{equation}
    \theta_{c} = 180^{\circ} - 2\sqrt{2\alpha^2},
    \label{eq:two}
\end{equation}
where $\theta_{ab}$ is the in-plane bond angle and $\theta_{c}$ is the out-of-plane bond angle.  In the L4S2 sample, we obtain $\theta_{ab} = 165.1 \pm 1.8^{\circ}$ and $\theta_{c} = 161.3 \pm 0.9^{\circ}$, while in the L2S2 sample, we obtain $\theta_{ab} = 168.2 \pm 1.2^{\circ}$ and $\theta_{c} = 164.4 \pm 0.6^{\circ}$.

  \begin{figure}
\includegraphics[width=2.3 in]{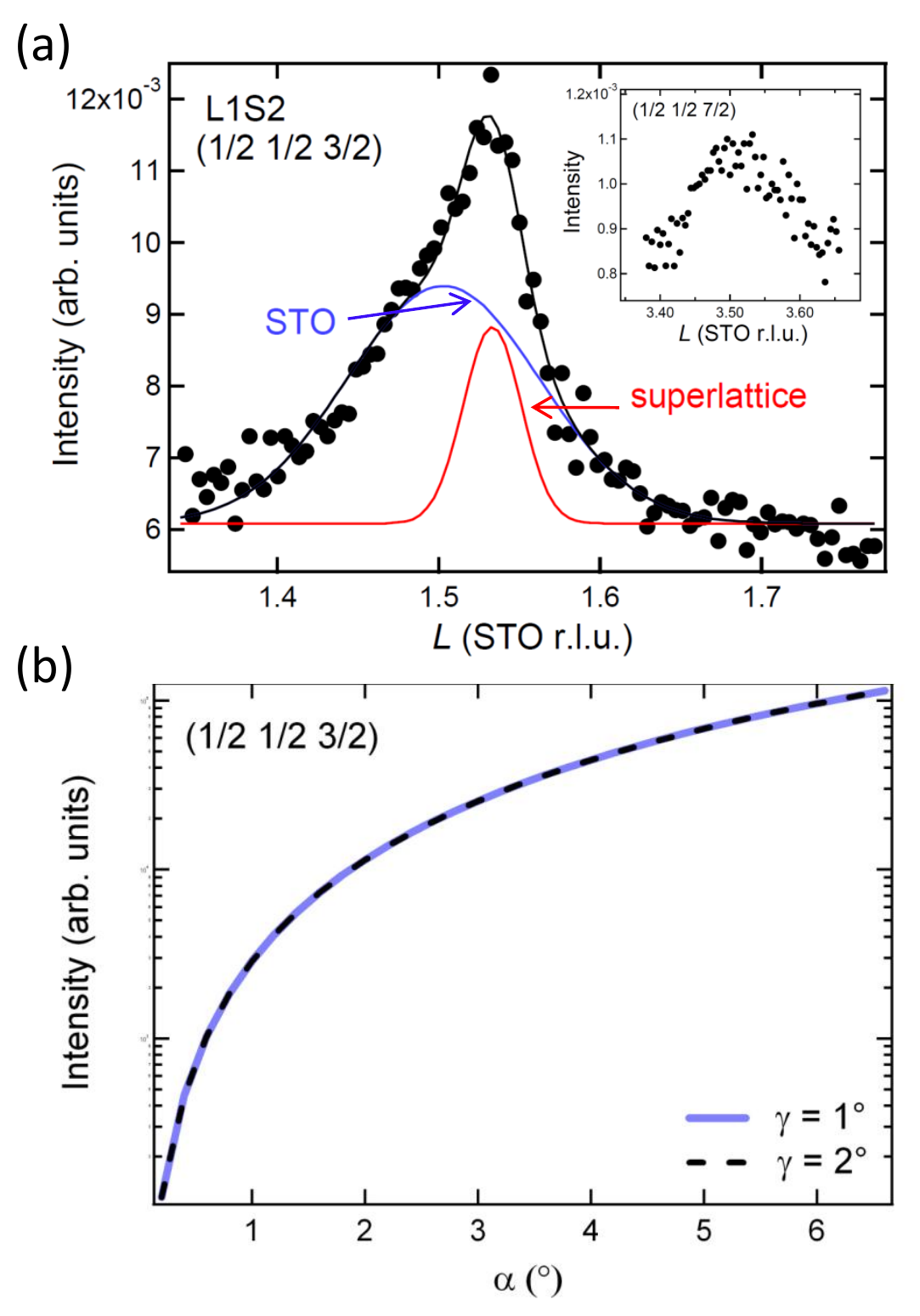}
\caption{(Color online) Scan through \textit{L} of the ($\nicefrac{1}{2}$ $\nicefrac{1}{2}$ $\nicefrac{3}{2}$) peak in the L1S2 superlattice (a).  The measured peak consists of a broad contribution from the STO substrate (curve centered at \textit{L} = 1.5) and a narrow contribution of the octahedral rotations within the superlattice (curve centered at \textit{L} = 1.533).  The substrate peaks obscure the weaker superlattice peaks, such as the ($\nicefrac{1}{2}$ $\nicefrac{1}{2}$ $\nicefrac{7}{2}$) (inset, a). The calculated intensity of the ($\nicefrac{1}{2}$ $\nicefrac{1}{2}$ $\nicefrac{3}{2}$) peak as a function of the $\alpha$ rotation angle (b).  At low $\gamma$ values, the ($\nicefrac{1}{2}$ $\nicefrac{1}{2}$ $\nicefrac{3}{2}$) intensity is independent of $\gamma$.}
\label{fig:L1S2}
\end{figure}

In the L1S2 sample, the half-order peaks are considerably less intense than those in the L2S2 and L4S2 samples and many peaks could not be accurately measured.  The ($\nicefrac{1}{2}$ $\nicefrac{1}{2}$ $\nicefrac{3}{2}$) peak measured from the L1S2 sample is shown in Fig.~\ref{fig:L1S2}(a).  In addition to the superlattice peak, there is a half-order peak from the STO substrate, which we speculate is defect related.  The substrate and superlattice contribution can be isolated as the substrate peaks are centered at exact half-integer values ($L$ = 1.5, 2.5, 3.5), while the superlattice peaks are shifted to higher momentum transfer ($L$ = 1.533, 2.566, 3.6).  Additionally, the substrate peaks are broad with a FWHM roughly 3 times larger than the superlattice half-order peaks.  The FWHM obtained from the L1S2 superlattice is roughly equal to that obtained from the L4S2 superlattice, indicating that the coherence length along the out-of-plane direction are equivalent in the two samples.

 \begin{figure}
\includegraphics[width=2.2 in]{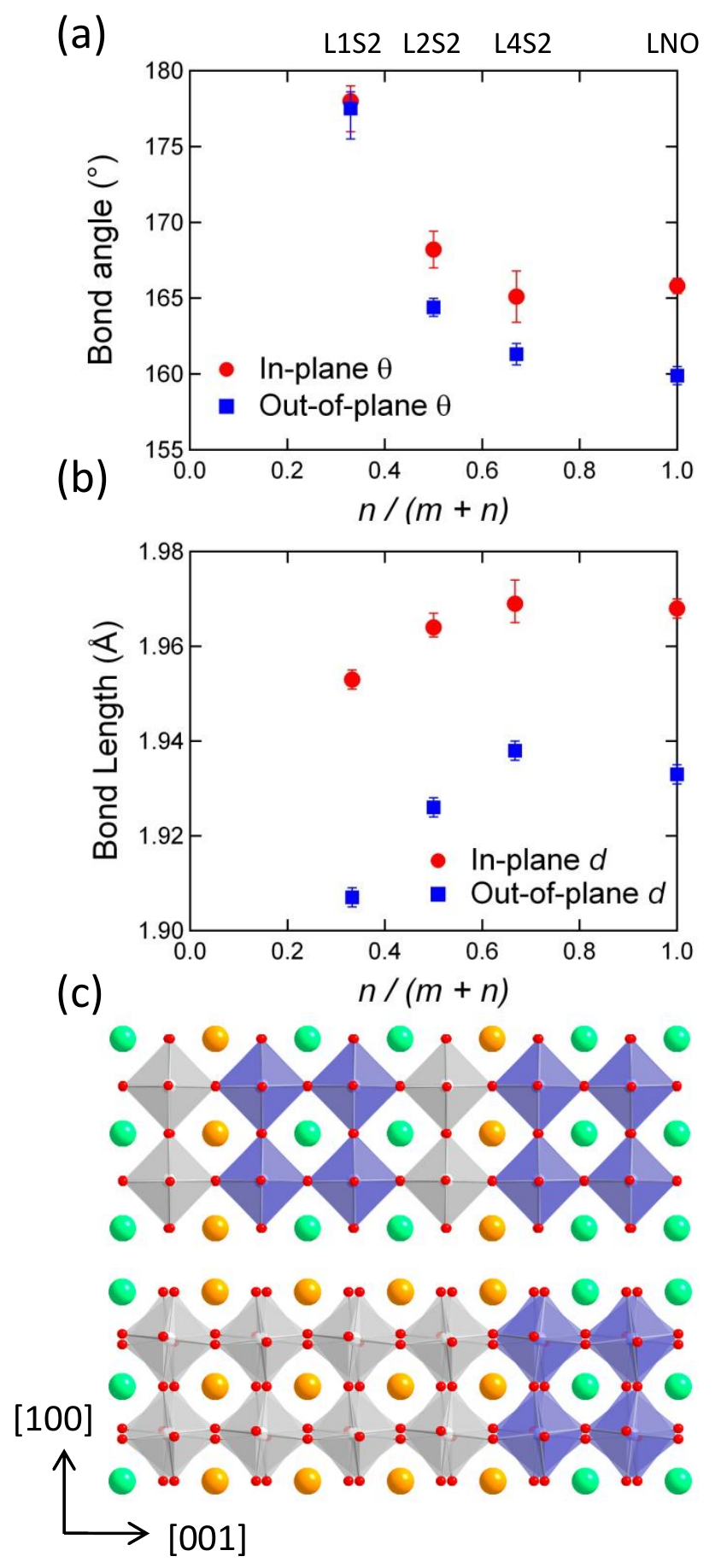}
\caption{(Color online) Bond angles (a) and bond lengths (b) presented as a function of superlattice composition.  The atomic structures of the L1S2 and L4S2 samples are shown in (c).  The top structure is the L1S2 superlattice, while the bottom structure is the L4S2 superlattice.}
\label{fig:angles}
\end{figure}

In order to analyze the L1S2 peak intensities, the $L$ scans taken at ($\nicefrac{1}{2}$ $\nicefrac{1}{2}$ $\nicefrac{3}{2}$) and ($\nicefrac{1}{2}$ $\nicefrac{3}{2}$ $\nicefrac{3}{2}$) were fit to two Gaussians, which were fixed at the substrate and superlattice positions.  The obtained intensities for the three samples are listed in Fig.~\ref{fig:table}(c).  These peaks are chosen as they are the two most intense peaks that we have measured and they are strongly dependent on $\alpha$ and $\gamma$.  The calculated intensity of the ($\nicefrac{1}{2}$ $\nicefrac{1}{2}$ $\nicefrac{3}{2}$) peak as a function of $\alpha$ is shown in Fig.~\ref{fig:L1S2}(b); the intensity changes by roughly 3 orders of magnitude from 0 to $6^{\circ}$ and is independent of $\gamma$ for small $\gamma$ angles.  By comparing the area under the ($\nicefrac{1}{2}$ $\nicefrac{1}{2}$ $\nicefrac{3}{2}$) peak with that measured in the L4S2 sample, a value of $\alpha = 0.9^{\circ}$ is obtained for L1S2.  Using the obtained $\alpha$ value, the ($\nicefrac{1}{2}$ $\nicefrac{3}{2}$ $\nicefrac{3}{2}$) intensities are compared for the L1S2 and L4S2 samples and $\gamma = 0.4^{\circ}$ is obtained for the L1S2 sample.  These $\alpha$ and $\gamma$ values correspond to $\theta_{ab} = 178^{\circ}$ and $\theta_{c} = 177.5^{\circ}$.  In this analysis, we assume that the in-plane coherence length of the L1S2 and L4S2 samples are approximately equal, as is observed for the out-of-plane coherence length.  As both the substrate and L1S2 ($\nicefrac{1}{2}$ $\nicefrac{3}{2}$ $\nicefrac{3}{2}$) peaks are centered about $H$ = 0.5, the in-plane coherence length of the L1S2 superlattice cannot be independently determined.  We have estimated the error bars for the L1S2 sample based on our uncertainty of the in-plane coherence length.  We note that we also applied this method of analysis to the L2S2 sample and obtained $\alpha = 5.0^{\circ}$ and $\gamma = 1.8^{\circ}$, in agreement with the result obtained from multiple peak fitting.

Given the rotation angles and lattice parameters ($a,c$), the average $B$-O bond lengths can be calculated from $d_{ab} = a/$(2 cos$\alpha$ cos$ \gamma$) and $d_c = c/$(2 cos$\alpha$ cos$ \alpha$).\cite{Glazer75}  The in-plane values ($d_{ab}$) are $1.969$, $1.964$, and $1.953$ \AA~in the L4S2, L2S2 and L1S2 samples, respectively, while the average out-of-plane values ($d_c$) are $1.938$, $1.926$, and $1.907$ \AA~in the L4S2, L2S2 and L1S2 samples, respectively.  We note that the $d_c$ values are the average of the Ni-O and Mn-O lengths and are expected to be different for the two octahedra due to the difference in local $c$-axis parameter between the LNO and SMO layers.

The average in-plane and out-of-plane bond angles and lengths for all three superlattices and a pure LNO film (from Ref. \onlinecite{May10}) are shown in Fig.~\ref{fig:angles}(a,b).  Both $\theta$ and $d$ are strongly dependent on superlattice composition.  The $\theta_{c}$ and $\theta_{ab}$ values can be tuned by $16^{\circ}$ and $13^{\circ}$, respectively, by altering the number LNO layers in each superlattice period.  This result illustrates the extent to which non-equilibrium octahedral behavior can be stabilized in short-period superlattices.  In this case, a nearly unrotated nickelate layer is present in the L1S2 superlattice, while the manganite layers in the L4S2 superlattice are highly rotated.  The corresponding atomic structures are shown in Fig.~\ref{fig:angles}(c).

Most importantly, the change in octahedral response to superlattice composition is fundamentally different from the octahedral behavior observed in single layer films as a function of epitaxial strain.  In compressively strained perovskites, in-plane bond lengths and angles are both reduced to accommodate the strain, while the opposite behavior is present under tensile strain.  As electronic bandwidth ($W$) is generally given a form such as $W = d^{-3.5}$cos$^2 \theta$, the strain-induced changes to bond angle and length act to alter $W$ in opposite manners, for instance with $d_{ab}$ and $\theta_{ab}$ acting to increase and decrease $W_{ab}$, respectively, in compressively strained films.\cite{Xie08}  In superlattices, compositional changes that increase $\theta$ also reduce $d$, and therefore both bond angles and length alter $W$ in the same manner.  Additionally, the range of in-plane bond angles enabled by the formation of short-period superlattices, here $13^{\circ}$, is much larger than that produced by epitaxial strain.  For instance, only a $2^{\circ}$ difference in $\theta_{ab}$ was observed between LaNiO$_3$ films under -1.1 to 1.7$\%$ strain.\cite{May10}  Therefore, the use of superlattice structures likely provides a more robust strategy than epitaxial strain in controlling bandwidth dependent properties in perovskites such as metal-insulator transitions, magnetic behavior, and optical band gaps.  We also suggest that changes in electrical conductivity found in perovskite superlattices compared to single layer films may arise in part due to superlattice-induced changes to the $B$O$_6$ octahedral behavior.\cite{Son10}

In conclusion, we have demonstrated how the $B$-O-$B$ bond angles and $B$-O bond lengths can be chemically controlled in complex oxide superlattices.  As the electronic structure is coupled to these bond angles and lengths, engineering the octahedral response of perovskites via the rational design of superlattice composition provides a novel framework for tuning metal-insulator transitions, ferroic properties, and band gaps in multifunctional oxides.

Work at Argonne was supported by the U.S. Department of Energy (DOE), Office of Basic Energy Sciences, under Contract No. DE-AC02-06CH11357.  Use of the Advanced Photon Source, an Office of Science User Facility operated for the DOE Office of Science by Argonne National Laboratory, was supported by the U.S. DOE under Contract No. DE-AC02-06CH11357. C.R.S is supported by the U.S. Department of Education under the GAANN program (\#P200A100134).  We are grateful to J. Rondinelli for useful discussions.


\begin{thebibliography}{24}
\bibitem[{not()}]{note}
\bibinfo{note}{The rotation angles about the [100] ($\alpha$) and [010]
  ($\beta$) axes are equal ($\alpha = \beta$) as the in-plane lattice
  parameters are equal ($a = b$) as described in Ref. \onlinecite{Glazer75}}.
\end{thebibliography}

\end{document}